\documentstyle[floats,aps,epsf]{revtex}
\begin{document}
\draft
\title{The two-phase approximation for black hole collisions:  Is it robust?}
\author{John Baker$^1$, Chun Biu Li$^2$\\
1. {\it Center for Gravitational Physics and Geometry, Department of 
Physics\\The Pennsylvania State University, 104 Davey Lab, University 
Park, PA 16802}\\
2. {\it Department of Physics, University of Utah, Salt Lake City, UT
84112}\\}

\maketitle
\begin{abstract}
Recently Abrahams and Cook devised a method of estimating the total
radiated energy resulting from collisions of distant black holes by
applying Newtonian evolution to the holes up to the point where a common
apparent horizon forms around the two black holes and subsequently
applying Schwarzschild perturbation techniques .  Despite the crudeness of
their method, their results for the case of head-on collisions were
surprisingly  accurate.  Here we take advantage of the simple radiated energy
formula devised in the close-slow approximation for black hole collisions
to test how strongly the Abrahams-Cook result depends on the choice of
moment when the method of evolution switches over from Newtonian to general
relativistic evolution.  We find that their result is robust, not depending
strongly on this choice.
\end{abstract}

\vspace{-8cm}
\begin{flushright}
\large CGPG-97/1-1\\
gr-qc/9701035\\
\end{flushright}
\vspace{7cm}
\bigskip

Black hole collisions are considered to be a likely source of
gravitational waves, expected to be detectable by the next generation
of gravitational wave antennas which will begin operation in a
few years.  Hence, there is currently strong motivation within the relativistic
astrophysics community to solve Einstein' equations for the case of two
colliding black holes.  The
full, numerical solution of Einstein's equations for a black hole
collision is a daunting problem, and it has yet to be achieved for any
realistic configuration.  It is thus useful to apply approximation
techniques wherever possible both to build intuition and as alternative
calculations against which the numerical results can be checked.  If
possible it would be desirable to have approximate estimates of how much
radiation will be generated by various types of collisions in order to identify
the most interesting problems for more detailed numerical study.

\begin{figure}[bt]
\caption{A cartoon illustration of how the two-phase approximation is constructed.}
\centerline{
\epsfysize=150pt
\epsfbox{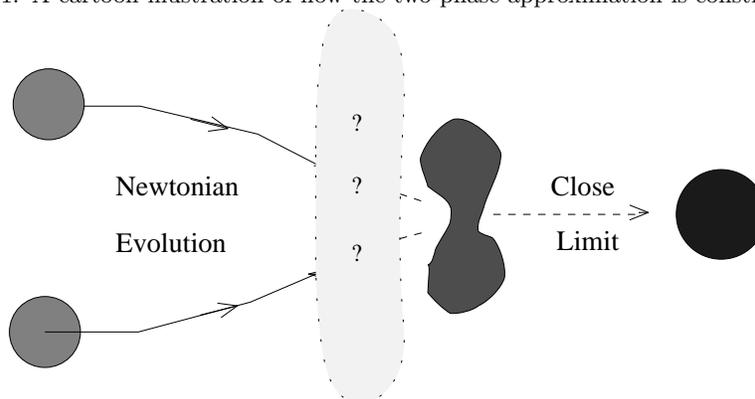}
}
\label{fig1}
\end{figure}

One useful  
approximation technique is the close-limit approximation. \cite{AP,PP} For
extremely close  black holes we can treat the system as a linear
perturbation of a single Schwarzschild black hole.  This approximation is
naturally expected to work well for evolution from initial data in which
the two initial black holes are 
sufficiently close to be well inside a common horizon, but for the cases
studied so far the results have shown good agreement with the numerical
calculations even to separations where no common horizon
exists.\cite{Anninos,Boost}  At the other extreme, if the black holes are
far apart and moving slowly we might apply the Newtonian approximation, whereby the holes are treated simply as point particles
moving in a flat background.  Clearly, we can never understand the later
and most important features of the collapse in only this far-limit.  On the
other hand,  Fig.\ 1 schematically  depicts the picture
which emerges upon considering these approximations together.  In a
``realistic'' head on
black hole collision the black holes are initially well-separated as is
indicated on the left side of the figure.  For this part of the in-fall we
can treat the holes as Newtonian point particles.  In the subsequent
evolution the black holes grow closer and their interaction grows stronger
until the two holes merge into a single distorted black hole.  This black hole
radiates away its distortions, ultimately leaving a Schwarzschild black
hole as the outcome.  Between the Newtonian region and the region where the
close-limit approximation applies there is a complicated transition period which
can only be understood by the exact numerical evolution.  
   However we may note that at least for axisymmetric collisions
nearly all of the radiation resulting from the collision is released in
the final moments of collapse, when the black holes have already merged
and the close-limit approximation works well.\cite{Boost}  One is then led to wonder if the
essential characteristics of the radiation can be captured by stretching
the  range of the Newtonian and close-limit approximations until they meet
to construct a two-phase approximation.
If we restrict
consideration to the total energy of the radiation generated by the
collapse, might we achieve a good estimate of this radiation by using
Newtonian mechanics to evolve the system of black holes until they are on
the verge of coalescence, and subsequently evolving from this point using
the close-limit approximation?

Abrahams and Cook first performed such a calculation in their study of
black holes with initial inward momentum.\cite{AC}  They applied the
close-limit 
approximation to nearby black holes with just enough inward momentum to
guarantee that they share a common apparent horizon in the initial
data reasoning that the horizon formation point was an identifiable
marker which roughly indicated the greatest separation at which the
close-limit calculation would be successful.  Then, they compared their results with those of the ``exact''
numerical calculation for the case of initially stationary black holes
released from such a separation that Newtonian in-fall would provide the
specified momentum at the moment of apparent horizon formation.   The
result was  excellent agreement with the numerical calculation.

 The Abrahams-Cook calculation  seems  to demonstrate the effectiveness of
two-phase approximate evolution for treating initially distant black hole
collisions.  However, let us suppose the numerical result didn't exist.
Could we then claim that the two phase approximation was generating a plausible
estimate of the radiation?  This is an important question because the
close-limit approximation is currently being applied to problems which have not
yet been solved numerically.  The critical question in determining the
usefulness of the two-phase approximation is whether the result is sensitive
to how the Newtonian first phase is joined to the close limit second phase.
For collapse from large separations, we know that the close-limit approximation
will wildly overestimate the radiation.  Similarly, we do not expect a
reasonable result if we allow the Newtonian phase of the evolution to
proceed too far.  Unfortunately, we are left to choose the point between
these extremes where we will switch from one type of evolution to the next. 
  Thus it may turn out that our result will be
strongly dependent on when we choose to make the transition.  Our method is
not useful if our uncertainty about the transition prevents us from making a
sufficiently precise estimate of the radiation.  Alternatively, if our
prediction for the radiation were very insensitive to our choice of
transition, then for a wide range of choices of transition, we would find
only a narrow range of energy predictions.  Then it would be compelling,
though not rigorously justified, to infer that the correct result lies in
that range.   The goal of the work
reported here is to determine whether there is such a narrow range of
energies resulting from a wide class of transition choices, or whether the
success of the Abrahams-Cook calculation was the result of a fortuitous
choice of a Newtonian to close-limit transition at the point of apparent
horizon formation.

Our method is straightforward.  We repeat the Abrahams-Cook
calculation\cite{AC} varying the point along the in-fall trajectory at which
Newtonian evolution ends and the close-limit evolution begins in order to
see how strongly the result depends on this choice.  A quick Newtonian-physics
calculation shows that for equal mass particles colliding head-on, 
\begin{equation}
	\left( \frac{l}{m} \right)_0 =\frac{l/m}{1-8(P/m)^2(l/m)},
\end{equation}
where $(l/m)_0$ is the initial separation from which the particles (black
holes) are released at rest,
$l/m$ is the separation at some later point, and $P/m$ is the magnitude of
each hole's momentum at that point. 
All quantities are scaled by $m$, the combined bare mass of the black holes.
The relation above defines a curve in the $l$-$P$ parameter space, the Newtonian
trajectory of a system of in-falling particles.  Thus, we have described
the Newtonian phase of the evolution.  At some point when the black
holes are close, we will halt the Newtonian evolution and begin
close-limit evolution.

  In principle, at this transition we will switch from a description of the
two black holes as a system of particles to a field-theoretic description,
where the fields 
obey an approximate form of Einstein's  equations.  This requires us
to specify values for the metric and its time-derivative on the initial
time-slice in terms 
of the parameters $l/m$ and $p/m$ which have resulted from the Newtonian
particle 
calculation.  In principle there is no unique way make this specification;
we must interpret the meaning of the particle parameters in the context of
initial field configuration.
In our case, to find the radiation, we will be
using the results of the ``close-slow'' approximation which has been
shown  to agree well with the numerical results for a broad range of
initial momenta when the black holes are nearly close enough to have
coalesced.\cite{Boost}   Implicitly, then, we will use the same description
of the initial data as has been used in the close-slow calculation wherein the 
data are specified  according to an approximation of
the procedure developed by Bowen and York \cite{BY} and numerically solved
for a wide variety of initial configurations by Cook.\cite{CookTh,Cook91}
In this formalism, the initial slice is assumed to 
be conformally flat with traceless extrinsic curvature.  A particular
family of solutions for the extrinsic curvature, labelled by $P$, is
chosen, and an elliptic boundary value problem determines the initial
metric.  After this, we must still  interpret the quantities $l$ and $m$ as
they will be realized for strongly interacting black holes.    While it is
natural to  take $l$ to be the proper separation
between the minimum area throats associated with each black hole, there
are several inequivalent ways to interpret the bare mass $m$.  We choose to
define  the bare mass of each black hole as the mass associated with the
area of its throat as is described in  Ref.\ \cite{Cook91}. 

  In practice, the second part of the evolution has already been solved.
The results of the close-slow calculation for the energy radiated
during the collision can be summarized in the following simple form.\cite{Boost},

\begin{equation}
\label{clsleq}
\frac{E}{M}=2.51\times10^{-2}\kappa_2^2(\mu_0)
-2.06\times10^{-2}\frac{{\rm coth}\,\mu_0\kappa_2(\mu_0)}{\Sigma_1}\left(
\frac{P}{M}
\right)
+5.37\times10^{-3}\left(
\frac{{\rm coth}\,\mu_0}{\Sigma_1}
\right)^2
\left(
\frac{P}{M}
\right)^2\ ,
\end{equation}
where $\mu_0$ is a parameter specifying the initial separation of the holes
and  both $\kappa_2$ and $\Sigma_1$ are known functions of $\mu_0$ described
in Ref.\ \cite{PP}.  Here all quantities are scaled by M, the ADM mass of the
spatial slice.  This equation gives us the total radiation directly in terms
of $\mu_0$ and $P/M$.
  
  However, to use the above result we need know $\mu_0$ and $P/M$ in terms
of the parameters $l/m$ and $P/m$.  Unfortunately, it is not natural
in the conformal formalism to calculate the initial data directly from
$l/m$ and $P/m$ as we interpret them here.  Only after calculating the
initial data sets in detail in terms of $\mu_0$ and $P$, and then
calculating the values of $l$, $m$ and $M$ 
can we ascertain their relationship among these parameters. Fortunately
Cook\cite{CookTh} has performed a large number of such calculations
generating tables of $l$, $m$ and $M$ in terms of $\mu_0$ and $P$ for many
cases covering the range 
of the parameter space with which we are interested.   Instead of 
repeating these calculations, we find that we are able to get a sufficient
estimate of $\mu_0$ and $P/M$ in terms of $l/m$ and $P/m$ by interpolation
from Cook's results.   Thus, we can apply the two-phase approximation to
the evolution of distantly in-falling black holes without really doing any
new calculations.  Newtonian methods give us a concise equation for the
first phase of evolution, the close-slow approximation gives us an
equation for the second phase, and the translation between the two methods
is accomplished by interpolation from existing studies of
axisymmetric initial data.

\begin{figure}[t]
\begin{minipage}[t]{8.5cm}
{
\epsfxsize=8.5cm
\epsfbox{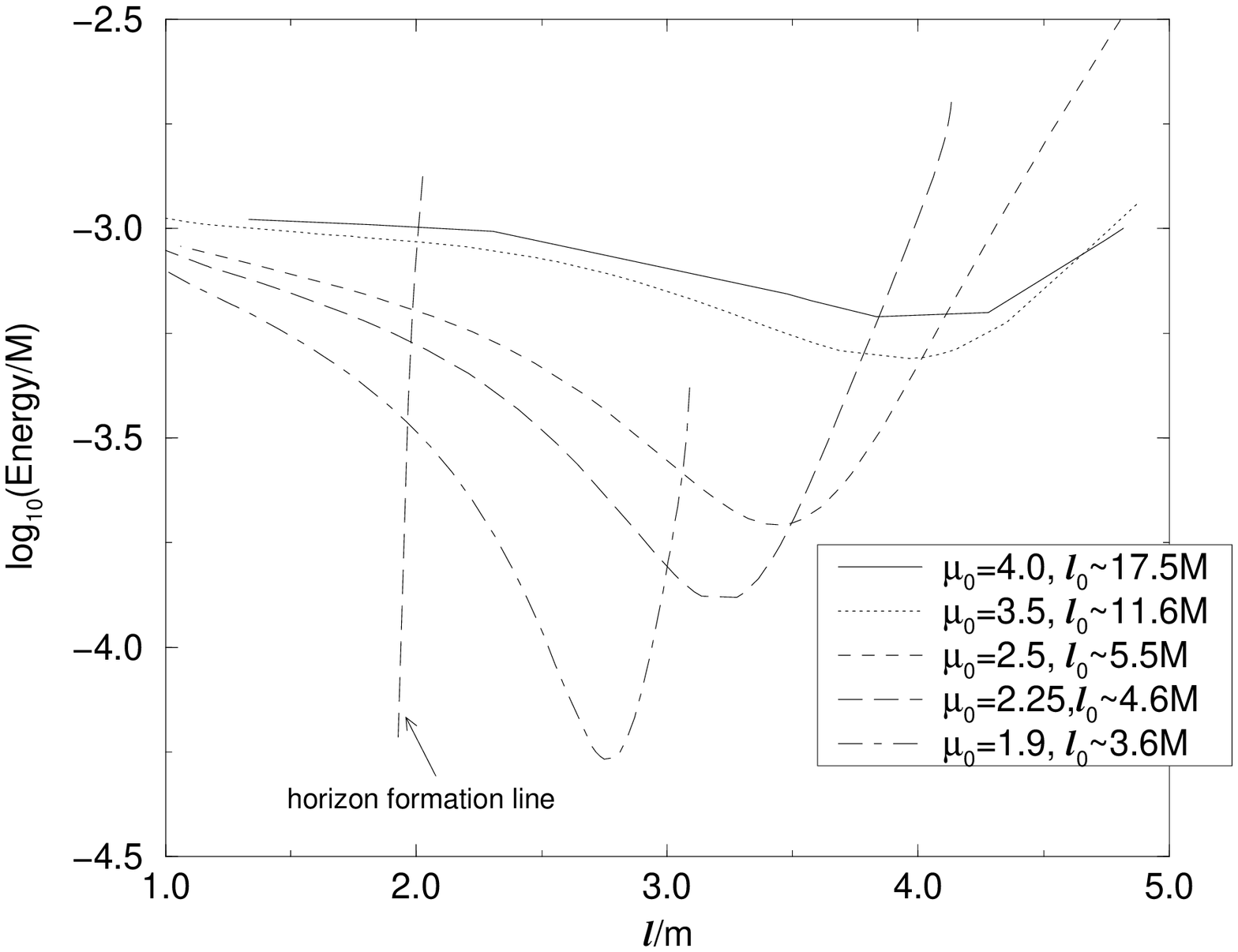}
\caption{Here we see that the amount of radiated energy is not sensitive to
the choice of transition.  The horizontal axis indicates the separation at
which we switch from Newtonian to close-limit evolution.  The  curves show
results for the radiation according to the two-phase approximation for
collisions of black holes released from several different initial
separations, $\mu_0$.  It is evident that for holes released from
significant initial distances the radiation result is not sensitive to the
choice of transition as long as the transition is made somewhere in the
vicinity of the separation at which a horizon first forms. }
\label{fig2}}
\end{minipage}
\hfill
\begin{minipage}[t]{8.5cm}
{
\epsfxsize=8.5cm
\epsfbox{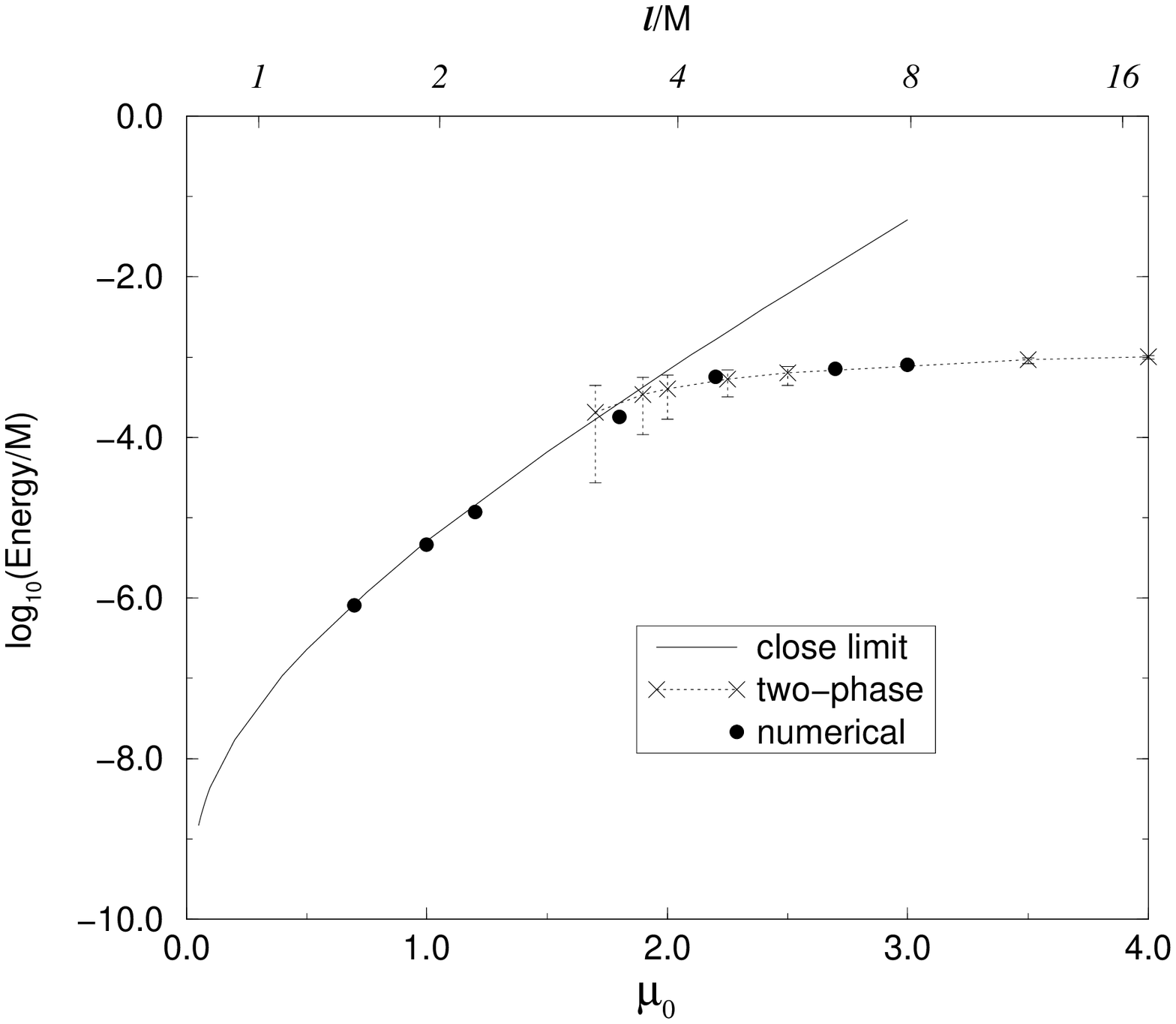}
\caption{This plot shows the excellent agreement of the two-phase
approximation with the numerical result.  The error estimates shown with the
two-phase data are based on our analysis of the sensitivity of these
results to the choice of separation at which we switch methods of
evolution.    }
\label{fig3}}
\end{minipage}
\end{figure}

We seek to determine the sensitivity of the two-phase approximation result
for the total radiation on the choice of transition.  In Fig. 2 we plot
this dependence for several different values of initial separation.  The
horizontal axis indicates the separation at which we make the transition
from Newtonian to close-limit  evolution.  For example, the curve labelled
$''\mu_0=2.5''$ shows that for holes released from rest at a separation of
$l \sim 5.5 M$   the two-phase technique will find a minimum amount
of radiation when the transition is made at $l = 3.5 m$.  The excellent agreement
with the numerical results initially discovered by Abrahams and Cook was
for the choice that the transition occurs at the point where the two black
holes are just close enough to share a common apparent horizon.  The
energies at the horizon formation line indicated in the figure thus
correspond to the Abrahams-Cook calculation.  Note that for 
large  initial separations, the total radiated
energy does not depend significantly on the choice of transition point in
the vicinity of the horizon formation line.  If the initial separation is
greater than $\mu_0 \sim 3.5$ is then even choosing to start the close-limit phase
when the black holes are more than twice as far apart only changes the
resulting radiation by less than a factor of two.  As the initial
separation becomes much closer the sensitivity begins to increase.  This
is not surprising since for initially close black holes there is
essentially no stage of the evolution 
where the Newtonian particle method is appropriate.  Fig. 3 shows a
comparison of the two-phase approximation with the Pullin-Price close-limit
result \cite{PP} and the result of the numerical calculation.\cite{num}
The error bars on the two-phase approximation result indicate the range of
energies obtained if we allow the evolution transition to occur at any proper
separation within $\frac{1}{2} m$ of the apparent horizon formation point.
As Abrahams and Cook have shown previously\cite{AC}, there is excellent
agreement with the numerical results.
  We conclude that the
Abrahams-Cook approximate result for the radiation generated by
in-fall from  large initial separation is insensitive to the choice of
transition from Newtonian to close-limit evolution.  Together with the fact
that the results for the radiation agree well with the numerical
calculation, this suggests that the the two-phase approximation method
may possibly be useful as a method for estimating the radiation generated
in other types of black hole collisions.

We would like to thank Richard Price for frequent guidance and
suggestions, and Andrew Abrahams, Greg Cook and Jorge Pullin for
many helpful comments. This work was supported in part by grants
NSF-PHY-9423950 and NSF-PHY-9507719.  We also acknowledge the support of
the Alfred P. Sloan foundation.

\end{document}